\documentclass[11pt]{iopart}
\jl{33}
\usepackage{iopams,graphicx}  
\begin{document}
\title[]{Spin and rotational symmetries in
unrestricted Hartree Fock states of quantum dots}
\author{ U. De Giovannini\dag, F. Cavaliere\dag, R. Cenni\ddag,  
M. Sassetti\dag, and B. Kramer\S}
\address{\dag Dipartimento di Fisica, Universit\`a di Genova, and CNR--LAMIA, Via Dodecaneso 33, 16146 Genova, Italy}

\address{\ddag Istituto Nazionale di Fisica Nucleare -- Sez. Genova\\Dipartimento di Fisica, Universit\`a di Genova, Via Dodecaneso 33, 16146 Genova, Italy}

\address{\S I. Institut f\"ur Theoretische Physik, Universit\"at
  Hamburg, Jungiusstra\ss{}e 9\\ 20355 Hamburg,
 and International University Bremen, Campus Ring 1, 28759 Bremen, Germany}

\ead{cavalier@fisica.unige.it}
\begin{abstract}
  Ground state energies are obtained using the unrestricted
  Hartree Fock method for up to four interacting electrons
  parabolically confined in a quantum dot subject to a magnetic field.
  Restoring spin and rotational symmetries we recover Hund's first
  rule.  With increasing magnetic field, crossovers between ground
  states with different quantum numbers are found for fixed electron
  number that are not reproduced by the unrestricted Hartree Fock
  approximation. These are consistent with the ones obtained with more refined techniques. We confirm the presence of a spin
  blockade due to a spin mismatch in the ground states of three and
  four electrons.
\end{abstract}
\pacno{73.21.La}
\maketitle
%
%
%
%
\section{Introduction}
\label{introduction}
In recent years, semiconductor quantum dots~\cite{kouwenhoven} have
been the subject of many experimental and theoretical investigations.
Their electronic properties can be controlled with a high accuracy by
applying external gate voltages and magnetic fields. Circular dots are
especially interesting. At low magnetic field $B$, the addition energy
of a two-dimensional circular dot has been found to exhibit pronounced
peaks for $N=2,6,12$ electrons~\cite{tarucha}. This suggests that
levels group themselves into shells, in analogy with atoms and nuclei and that
closed-shell configurations are particularly stable. The behaviour of
open-shell states has been found to be consistent with Hund's first
rule~\cite{baym}. At non--zero magnetic field, transitions between
ground states involving total spins and angular momenta have been
observed~\cite{mceuen,haug,ciorga}. These have a profound effect on
the transport properties of a quantum dot. For instance, if
the total spins of two ground states with $N$ and $N+1$ electrons
differ by more than $\hbar /2$ spin blockade of electron transport is
predicted~\cite{weinman}. This has been recently observed in the
conductance of quantum dots~\cite{weinman2}.

Theoretically, the electronic structure of quantum dots has been
studied using many different techniques~\cite{reimann}. Exact
diagonalization
(ED)~\cite{dineykhan,merkt,mikhailov2,mikhailov,wojs,szafran,maksymB,kyriakidis}, configuration interaction (CI)~\cite{wensauer,rontani}, stochastic variational method~\cite{varga} and the pocket state method~\cite{hausler} allow to calculate ground and excited state energies and their quantum
numbers with very good accuracy. Also quantum Monte Carlo methods~\cite{pederiva,pederiva2,egger,harju,bolton} have been employed: they provide accurate estimates for ground and excited states energies, although total spin symmetry is not always preserved~\cite{egger,harju}. By means of all these
techniques, shell structure and Hund's rule have been analyzed in
detail. In addition, ``magic'' values for the total angular momentum
have been predicted to occur for dot states with a given total spin.
They occur if electrons in a quantum dot strongly interact and arrange
themselves in a rotating Wigner
molecule~\cite{mikhailov,harju,maksym,ruan,filinov,bao}. All the above methods are computationally very expensive and can be used for relatively low electron numbers $N\leq13$: only with Monte Carlo methods electron numbers up to $N=24$ have been reached~\cite{pederiva2} at zero magnetic field.

Other methods are used for treating systems with larger electron
numbers, like the Hartree Fock approach (HF)~\cite{landmanprl,landman3,reusch0,reusch,lipparini,hawrylak1,hawrylak2}
and density functional theory~\cite{manninen,hirose,gattobigio,harju2}. They generally provide less accurate estimates of the ground state energy and the resulting wave functions can have unphysically broken symmetries.
The HF methods are paradigmatic, in that the variational ground state wave function is a single Slater determinant which does not properly include correlations and in general is not an eigenfunction of the total spin~\cite{szabo}. In contrast with space restricted HF (RHF) methods, space {\em unrestricted} HF (UHF) methods~\cite{landmanprl,reusch0} systematically allow for symmetry breaking allowing, as a starting point for the calculations, wave functions without rotational
invariance even in the case of circularly symmetric dots. While UHF
yields a lower estimate for the ground state energy, it can have
severe drawbacks when considering physical properties of the ground
state like the total spin.  For instance, UHF
calculations sometimes fails predicting Hund's rule, in contradiction to
experiments, and in contrast to results obtained with other methods. In addition, wave functions with broken symmetries do
not allow to determine the total spin and the angular momentum.
Methods to restore the correct symmetries have been pioneered in the
60th of the last century~\cite{loewdin1,ring,loewdin2} and were used
in the context of quantum dots, mostly for the rotational
symmetry~\cite{landman2,landman4,landman5,koonin,YL2002}. Restoration of the
{\em spin} symmetry in quantum dots has received much less
attention~\cite{landman2}.

In this paper, we apply a systematic projection procedure to restore
both the {\em spin} and the {\em rotational} symmetries of ground state
variational wave functions obtained by UHF calculations for quantum
dots with up to four electrons, including a magnetic field. It is our
aim to show that, after {\em all the symmetries} are restored, the
wave functions are considerably improved and show several physical
features which are not reproduced by the straightforwardly applied UHF
method. Restoring symmetries introduces correlations, absent within the single UHF Slater determinant, leading to better energy estimates.

We demonstrate the efficiency and accuracy of the projection procedure by
comparing our results with those of the methods mentioned
above. The main findings are: (i) at zero magnetic field, the
first Hund's rule is recovered for four
electrons which has been claimed earlier to be violated by UHF \cite{landmanprl}; 
(ii) for nonzero magnetic field, many crossovers between ground states
with different total spins and angular momenta of up to four electrons
are found that are completely missed by using the HF method alone.

The paper is organized as follows. In section~\ref{theham}, the UHF
procedure is briefly sketched, emphasizing the role of broken
symmetries. In section~\ref{secproj}, the projection techniques
employed in the symmetry restoration are discussed. Results for zero
and non-zero magnetic field are discussed in section~\ref{results}. The
paper is concluded by pointing out the perspectives for obtaining results for higher numbers of electrons that are not accessible by other methods.
%
%
%
%
\section{The Unrestricted Hartree Fock Approximation}
\label{theham}
The Hamiltonian for a system of $N$ interacting electrons,
parabolically confined in the $x$--$y$ plane and subject to a perpendicular
magnetic field ${\bf B}=B{\bf e}_{z}$ (${\bf e}_{z}$ the $z$ axis unit vector, here and in the following $\hbar=c=1$) is
\begin{equation}
\label{hamil}
{H}=\sum_{i=1}^{N}{H}_{0}({\bf
r}_{i},{\bf p}_{i},s_{zi})+\sum_{i>j}{V}({\bf r}_{i}-{\bf r}_{j})
\end{equation}
where
\begin{equation}
\label{hamiltonian}
{H}_{0}({\bf r},{\bf p},s_{z})=\frac{{\left[{\bf p}+e{\bf A}({\bf
    r})\right]}^{2}}{2m^{*}}+\frac{m^{*}\omega_{0}^{2}}{2}{\bf
r}^{2}+g^{*}\mu_{B} B s_{z}\,,
\end{equation}
with Coulomb interaction potential ${V}({\bf
  r})={e^2}/{4\pi\varepsilon_{0}\varepsilon}{|{\bf r}|}$, ${\bf
  B}={\rm rot} {\bf A}$, effective electron mass $m^{*}$,
confinement frequency $\omega_{0}$,  effective $g$--factor $g^{*}$ and the
Bohr magneton $\mu_{B}$. The $z$ component of the
$i$--th spin is $s_{zi}=\pm 1/2$. Furthermore, $-e$ is the
electron charge and $\varepsilon_{0}$ ($\varepsilon$) the vacuum
(relative) dielectric constant. The Hamiltonian~(\ref{hamil}) commutes
with the total angular momentum
${L}={l}_{1}+\ldots+{l}_{N}$ ($z$ component in two
dimensions), the $z$-component of the total spin
${S}_{z}={s}_{z 1}+\ldots+{s}_{z N}$ and the total spin
${\bf S}={\bf s}_{1}+\ldots+{\bf s}_{N}$. The single-particle
term ${H}_{0}$ is exactly diagonalizable. It yields the Fock-Darwin
(FD) spectrum
\begin{equation} 
\label{fdspect}
\epsilon_{n,m,s_{z}}=\Omega\left(2n+|m|+1\right)+
\frac{\omega_{c}}{2}m+g^{*}\mu_{B} B s_{z}\, ,
\end{equation}
with the corresponding harmonic eigenfunctions $\phi_{n,m,s_{z}}({\bf
  r})$~\cite{fd}. Here, $n$ and $m$  are principal and angular
momentum quantum numbers. Further, we introduced the
cyclotron frequency $\omega_{c}=eB/m^{*}$, and the effective
confinement frequency $\Omega=(\omega_{0}^{2}+\omega_{c}^{2}/4)^{1/2}$.

When including interactions, the problem is in general not solvable.
Here, we deal with the electron interactions by using as a starting point the HF
approximation. The interacting, many--body ground state wave function
is written as a single Slater determinant consisting of $N$ orbitals of
the form $\psi^{s_{z}}({\bf r})$, assumed to be eigenfunctions of
${s}_{z}$
\begin{equation}
\psi_{i}^{s_{z}}({\bf r})=a_{i}({\bf r}){\alpha}\delta_{s_{z},+1/2}+b_{i}({\bf
  r}){\beta}\delta_{s_{z},-1/2}\, ,
\end{equation}
with $a,b$ denoting the spatial parts, and ${\alpha}$, ${\beta}$
the spinors corresponding to $s_{z}=\pm 1/2$, respectively. Denoting $N_{+}$ and $N_{-}$ the number of spin up and spin down orbitals,
respectively, the Slater determinant is
\begin{equation}\label{det}
\left|\Psi^{S_{z}}\right>={(N!)}^{-1/2}
{\rm det}\{a_{1}{\alpha},\ldots,a_{N_{+}}{\alpha},b_{1}{\beta},\ldots,b_{N_{-}}{\beta}\}\, ,
\end{equation}
eigenfunction of $S_{z}$. Slater determinants are not in general eigenfunctions of ${S}^2$, unless $|S_{z}|=N/2$ when the HF solution has total spin quantum number $S=N/2$. Thus, HF solutions with $S_{z}<N/2$ are not consistent with the spin symmetry of the Hamiltonian. This is known as {\em spin contamination}~\cite{szabo}: a
single Slater determinant is in general a superposition of many
eigenfunctions with different total spins but with a given $S_{z}$.

The determinant is variationally optimized in order to minimize the energy
\begin{equation}
E^{S_{z}}=\frac{\left<\Psi^{S_{z}}\right|{H}\left|\Psi^{S_{z}}\right>}
{\left<\Psi^{S_{z}}\right|\left.\Psi^{S_{z}}\right>}\, ,
\end{equation} 
giving rise to the well known $N$ coupled integro-differential equations for the 
orbitals $\psi_{i}^{s_{z}}({\bf r})$,
\begin{eqnarray}
&\left[{H}_{0}({\bf r})+\int{\rm d}{\bf
    r}' \rho({\bf r}'){V}({\bf r}-{\bf r}')\right]\psi_{i}^{s_{z}}({\bf
r})&\nonumber\\&-\sum_{j=1}^{N_{s_{z}}}\left[\int{\rm d}{\bf
  r}'\psi_{j}^{s_{z} *}({\bf
    r}')\psi_{i}^{s_{z}}({\bf
    r}') {V}({\bf r}-{\bf r}')\right]\psi_{j}^{s_{z}}({\bf
  r})=E_{i}^{s_{z}}\psi_{i}^{s_{z}}({\bf r})\, &\label{HFE}
\end{eqnarray}
where $N_{s_{z}}=N_{+}\delta_{s_{z},1/2}+N_{-}\delta_{s_{z},-1/2}$ and the electron density is 
\begin{equation}
\rho({\bf r})=\sum_{s_{z}=\pm 1/2}\sum_{j=1}^{N_{s_{z}}}\left|\psi_{j}^{s_{z}}({\bf
    r})\right|^2\, .
\end{equation}
Within a given sector with fixed $S_{z}$, equation~(\ref{HFE}) is solved self--consistently, starting from an
initial guess for the orbitals. {\em Spatially unrestricted}
initial guesses are used: orbitals are assumed such that they are not eigenfunctions of
the angular momentum~\cite{landmanprl,landman2,reusch0} (UHF method). With
this, solutions of the HF equations corresponding to non--rotationally
invariant electron densities, are obtained. These symmetry--broken
solutions enhance the number of variational degrees of freedom, thus
generally leading to better energy estimates. Therefore, in addition to the lack of total spin symmetry already present in RHF calculations, UHF solutions do not possess the spatial symmetry of~(\ref{hamil}). Their predictive power for total spin and angular momentum of
the ground state is therefore limited. As a consequence, the symmetries of UHF wave functions must be suitably adjusted in order to address these properties.

It is important to note that for given $N$ and $S_{z}$, many local minima of the energy surface
 can be found depending on the initial guess used for the UHF calculation. In general,
one finds a sequence of states $|\Psi_{k}^{S_{z}}\rangle$ ($k=1,2,\ldots$) with
energies $E_{1}^{S_{z}}<E_{2}^{S_{z}}<\ldots$.  For a given $S_{z}$,
one has to perform extensive scans over the space of initial
conditions in order to achieve confidence that the first state in the
above sequence is the best UHF state, with the lowest attainable
energy. The {\em UHF ground state} is then defined as the state with
the lowest among the energies $E_{1}^{S_{z}}$. This also determines the value of $S_{z}$ for the UHF ground state.
%
%
%
%
\section{Restoring Symmetries}
\label{secproj}
In order to reflect the symmetries of the Hamiltonian in the UHF
solutions, projection operators~\cite{loewdin1,ring} can be applied to
the wave functions. To avoid confusion, in this section operators are denoted by a over hat. Consider $\hat{P}_{L}$ and $\hat{P}_{S}^{S_z}$ which project the UHF solution
onto subspaces with well-defined $L$ and $S$. They satisfy the
commutation rules $[\hat{P}_{S}^{S_z},\hat{P}_{L}]=[\hat{P}_{S}^{S_z},\hat{H}]=[\hat{P}_{L},\hat{H}]=0$. For a given UHF solution $|\Psi^{S_z}\rangle$ we define the projected state with spin and angular momenta by $|\Psi_{L,S}^{S_z}\rangle=\hat{P}_{L}\hat{P}^{S_z}_{S}|\Psi^{S_z}\rangle$. The corresponding energy is
\begin{equation}
\label{energyp}
E_{L,S}^{S_{z}}=\frac{\langle\Psi_{L,S}^{S_z}|\hat{H}|\Psi_{L,S}^{S_{z}}\rangle}
{\langle\Psi_{L,S}^{S_z}|\Psi_{L,S}^{S_{z}}\rangle}=
\frac{\langle\Psi^{S_z}|\hat{H}|\Psi_{L,S}^{S_{z}}\rangle}
{\langle\Psi^{S_z}|\Psi_{L,S}^{S_{z}}\rangle}\, ,
\end{equation}
where we used the commutation rules presented above and the idempotency of projection operators.
The spin projector $\hat{P}_{S}^{S_{z}}$ can be written as~\cite{loewdin1}
\begin{equation}
\label{projs}
\hat{P}_{S}^{S_z}=\prod_{k=|S_{z}|,k\neq S}^{N/2}\frac{\hat{S}^{2}-k(k+1)}{S(S+1)-k(k+1)}\ .
\end{equation}
Its action on the UHF wave function is~\cite{loewdin1,projS1} 
\begin{equation}
\label{project1}
\hat{P}_{S}^{S_z}|\Psi^{S_z}\rangle=\sum_{q=0}^{N_{<}}C_{q}(S,S_{z},N,N_+)|T_{q}\rangle
\end{equation}
where $N_{<}={\rm min}\{N_{+},N_{-}\}$ and $C_{q}(S,S_{z},N,N_+)$
\begin{equation}\label{sanibel}
C_{q}(S,S_{z},N,N_+)=\frac{2S+1}{1+N/2+S}\sum_{k=0}^{S-S_z}(-1)^{q+S-S_z-k}\frac{{S-S_z\choose k}{S+S_z\choose S-S_z-k}}
{{N/2+S\choose N_+-q+k}} 
\end{equation}
are the Sanibel coefficients~\cite{projS1,ruitz}. The term
$|T_{q}\rangle=|T_{q}^{(1)}\rangle+\ldots+|T_{q}^{(n_q)}\rangle$ is the
sum of all of the
\begin{equation}
\label{numberofterms}
n_q={N_{+} \choose q}{N_{-} \choose q}
\end{equation}
distinct Slater determinants obtained by interchanging in the initial
determinant $\mid \Psi ^{S_z}\rangle$ {\em all} $q$ spinor pairs with
opposite spins.

The projection operator on the total angular momentum $L$ is~\cite{ring}
\begin{equation}
\label{projl}
\hat{P}_{L}=(2\pi)^{-1}\int_{0}^{2\pi}{\rm d}\gamma\ e^{-iL\gamma}e^{i\hat{L}\gamma}\, .
\end{equation}
Acting on $|T_{q}\rangle$ with $\exp{(i\hat{L}\gamma)}$ results in
$|T_{q}(\gamma)\rangle$, a determinant where all the orbitals are
rotated by $\gamma$ around the $z$ axis. Combining~(\ref{project1})
and~(\ref{projl}) we have
\begin{equation}
\label{project2}
|\Psi_{L,S}^{S_{z}}\rangle=(2\pi)^{-1}\sum_{q=0}^{N_{<}}C_{q}(S,S_{z},N,N+)\int_{0}^{2\pi}{\rm
  d}\gamma\ e^{-iL\gamma}|T_{q}(\gamma)\rangle\, .
\end{equation} 
The projected state~(\ref{project2}) is a sum of many Slater
determinants. This indicates that a high degree of correlation has
been introduced by applying the projection technique to the UHF
scheme. As is clear from (\ref{project2}), spin projection implies
calculations which involve many Slater determinants. Therefore the
second form of~(\ref{energyp}) is especially useful from the
computational point of view.  A detailed description of the procedure
is developed in \ref{toy} for $N=3$.

We have implemented numerically the evaluation of~(\ref{energyp}),
using~(\ref{project2}) and standard theorems for the evaluation of the
Hamiltonian matrix elements~\cite{loewdin2}. In this work, we will
determine the {\em projected ground state}, defined as the {\em
projected state with the lowest energy}. This procedure, labelled
``projected Hartree Fock'' (PHF) in the following, is far from being
trivial: in particular, it is {\em not} sufficient to project the UHF
ground state alone to determine the PHF ground state. If several UHF
minima $E_{i}^{S_{z}}$ are almost degenerate, all the corresponding
$|\Psi_{i}^{S_{z}}\rangle$ must be projected. For each
$|\Psi_{i}^{S_{z}}\rangle$, the projected energies $E_{i,L,S}^{S_{z}}$
attain a minimum
$\bar{E}_{i,\bar{L}_{i},\bar{S}_{i}}^{S_{z}}<E_{i}^{S_{z}}$ for given
$\bar{L}_{i},\bar{S}_{i}$. The PHF ground state is thus the lowest
among $\bar{E}_{i,\bar{L}_{i},\bar{S}_{i}}^{S_{z}}$. 

For a given UHF solution (corresponding to fixed $B$ and $S_z$), the spin projection involves the evaluation of all the
overlapping matrix elements $\langle T_{0}|\hat{H}|T_{q}^{(i)}(\gamma)\rangle$ (see \ref{toy} for details).  
This constitutes a computational overhead, especially for high particle numbers.
For given  $S_z$ and $S$, the number of matrix elements to be evaluated is given by $n_{S_z}=\sum_{q=0}^{N_<}n_q$.
The worst case scenario occurs for the minimal values of $S_z$. 
 As an example, for even $N$, the number of matrix elements is
\begin{equation}
n_{S_z=0}=\frac{2^N}{\sqrt{\pi}}\frac{\Gamma((1+N)/2)}{\Gamma(1+N/2)}\stackrel{N\to\infty}{\longrightarrow}\frac{2^{N}}{\sqrt{N}}\, ,
\end{equation}
(a similar formula with the same asymptotic behavior holds for odd
$N$).  Numerically performing the angular momentum projection requires
the discretization of $\gamma\in[0,2\pi]$ in $n_{L}$ points, followed
by a fast Fourier transform (FFT) procedure. The lower bound for the
latter is determined by the maximum angular momentum $|L|\leq L_{max}$
desired, independent of $N$.  We point out that, once the sampling has
been performed, the FFT simultaneously produces all the angular
momenta projections for $|L|\leq L_{max}$.  We have checked that, in
order to achieve good convergence with $L_{max}=20$ (much higher than
the highest $|L|$ involved in the results discussed below), one needs
$n_{L}=256$. Thus, for the case $N=4$, $S_{z}=0$, fixed $S$ and
$|L|\leq 20$, we need to evaluate $n_{S_z} n_L=1536$ matrix
elements. This compares favorably with respect to more refined
methods: for instance, exact diagonalization for $N=4$, $S_z=0$,
$S=2$, $L=14$, needs up to 19774 Slater
determinants~\cite{szafran}. Also for increasing numbers of electrons
PHF performs well: in the case $N=6$, $S_{z}=0$, $S=0$, $L=0$ (not
discussed in the present work), CI needs 661300 configurational states
functions (linear combination of Slater determinants)~\cite{rontani}
while our procedure requires 5120 matrix elements to evaluate not only
$L=0$ but all $|L|\leq 20$. We expect a similar performace gain for
even larger number of electrons.

In the next section we will show that the PHF ground state has lower
energy than UHF one. Most importantly, all quantum numbers of the
projected ground state can be determined, in contrast to UHF.
%
%
%
%
\section{Results}
\label{results}
Before presenting our results, we provide some technical details of
the numerical method. We rewrite equations~(\ref{HFE}) by employing
the non interacting FD-basis and obtain a nonlinear Pople-Nesbet
eigenvalue problem~\cite{szabo}. For each value of $B$, the 75
lowest-lying FD states per spin direction have been assumed to represent the basis. This
guarantees a fair convergence of the UHF procedure for $N\leq 4$. By
comparing results with those obtained with a smaller basis set (55 elements), the relative uncertainty
of ground state energies has been assured to be $\sim10^{-6}$ in
all of the data shown below. For a given number of electrons, many
different UHF solutions for all possible values of $S_{z}$ have been
determined. Subsequently, the projection procedure has been
implemented, taking particular care of the cases with almost degenerate
UHF states as discussed above.
%
%
%
%
\subsection{Zero magnetic field}
\label{zerofield}
At zero magnetic field, expressing energies in units $\omega_{0}$ and lengths in units $\ell_{0}={(m^{*}\omega_{0})}^{-1/2}$, the Hamiltonian~(\ref{hamil}) depends only upon the dimensionless parameter
\begin{equation}
\label{lambda}
 \lambda=\frac{e^{2}}{4\pi\varepsilon_{0}\varepsilon\ell_{0}\omega_{0}}\, ,
\end{equation}
which represents the relative strength of the interaction.  
\begin{table}[htbp]
\begin{center}
\begin{tabular}[htbp]{|c||l|l|l|}
\hline
{\em N}&{$E_{\rm UHF}$}&{$E_{\rm PHF}$}&{$E_{\rm DMC}$}\\
\hline \hline
{ 2}&3.956 ($S_{z}=0$)&3.816 (L=0, S=0)&3.649 (L=0, S=0)\\
\hline
{ 3}&8.236 ($S_{z}=3/2$)&8.155 (L=1, S=1/2)&7.978 (L=1, S=1/2)\\
\hline
{ 4}&13.786 ($S_{z}=0$)&13.554 (L=0, S=1)&13.266 (L=0, S=1)\\
\hline
\end{tabular}
\caption{Comparison of the ground state energies
  and quantum numbers for a dot with $N=2,3,4$ electrons,
  calculated with three different methods. Column 2: UHF ground
  state. Column 3: PHF ground state. Column 4: DMC ground
  state, extracted from~\cite{pederiva}. In columns 3 and 4 $S_{z}$ is not
  indicated since at zero magnetic field spin multiplets are degenerate.
  Interaction parameter $\lambda=1.89$ (see Eq.~(\ref{lambda})). Energies in units $\omega_{0}$. }
\label{table1}
\end{center}
\end{table}
Table~\ref{table1} shows ground state energies for $N=2,3,4$ electrons
with $\lambda=1.89$, chosen in order to compare our results with
those obtained with diffusion Monte Carlo (DMC)~\cite{pederiva} (third
column). The second column shows the UHF energy, the third column the
PHF ground state energy while the fourth DMC data. Since at zero
magnetic field the spin multiplets are degenerate with respect to
$S_{z}$, the latter is not mentioned in columns 3, 4.

As already discussed, projection improves the ground state energy. The quantum numbers for the PHF ground state agree with the ones obtained with DMC. The computed energies differ for at most about 4\% ($N=2$ case) and improve increasing $N$.
Our results agree also with other published data: for $N=2$, a singlet ground state has also been found by means of ED~\cite{dineykhan,merkt,pfannkuche}. For $N=3,4$ electrons, the quantum numbers predicted by PHF agree with those predicted by the CI method~\cite{wensauer,rontani} and the path integral Monte Carlo method (PIMC)~\cite{egger}.

The ground state for $N=4$ is a triplet, in agreement with Hund's
first rule. Previously published calculations~\cite{landmanprl} found
that the UHF ground state for four electrons has $S_{z}=0$. Our UHF
calculation, although performed with a confinement energy different
from that used in~\cite{landmanprl}, confirms this result
(table~\ref{table1}, column 2). A ground state with $S_{z}=0$
previously has been interpreted~\cite{landmanprl} as violating Hund's
first rule.  However, the latter deals with the {\em total spin} of
the electrons, an undefined quantity in UHF solutions.
Projecting the total spin allows to find that the ground
state has $S=1$ which, due to the degeneracy of the spin multiplets,
is compatible with $S_{z}=0$.

\begin{table}[htbp]
\begin{center}
\begin{tabular}[htbp]{|c||l|l|l|}
\hline
{$\lambda$}&{$E_{\rm UHF}$}&{$E_{\rm PHF}$}&{$E_{\rm CI}$}\\
\hline \hline
2&14.140 ($S_{z}=0$)&13.899 (L=0, S=1)&13.626 (L=0, S=1)\\
\hline
4&19.581 ($S_{z}=2$)&19.330 (L=0, S=1)&19.035 (L=0, S=1)\\
\hline
6&24.139 ($S_{z}=2$)&23.880 (L=0, S=1)&23.598 (L=0, S=1)\\
\hline
8&28.272 ($S_{z}=2$)&27.993 (L=0, S=1)&27.671 (L=0, S=1)\\
\hline
\end{tabular}
\caption{Ground state energies and quantum numbers for the $N=4$ case at different interaction parameter
  $\lambda$ (see Eq.~(\ref{lambda})). Column 2: UHF ground state. Column 3: PHF ground state. Column 4: CI ground state, extracted from~\cite{rontani}.  In
  columns 3 and 4 $S_{z}$ is not indicated since for $B=0$ spin
  multiplets are degenerated. Energies in units $\omega_{0}$.}
\label{table2}
\end{center}
\end{table}

To probe the spin properties of the $N=4$ case, we study it for increasing $\lambda$. For strong interactions, the UHF method is commonly assumed to favor spin polarized states~\cite{pfannkuche}. Thus, a crossover of the UHF ground state to $S_{z}=2$ is expected. Table~\ref{table2} shows a comparison of UHF, PHF and CI~\cite{rontani} results. The UHF ground state for $\lambda\geq 4$ has $S_{z}=2$.  This is not compatible with a triplet state, and  indicates a violation of Hund's first rule. The violation is striking since $S_{z}=2$ is an uncontaminated $S=2$ state. On the other hand, as shown in column 3, for $\lambda\geq 2$ PHF predicts a {\em triplet} ground state, consistent with Hund's rule and in qualitative agreement with the CI results (column 4). For $\lambda\geq 6$, the PHF ground state originates from a low lying excited state (local minimum) of UHF (see below). When increasing the strength of the interaction, it is expected that correlations, measured by
the difference between the energies of the ground state obtained by the RHF approximation and the exact ground state, become increasingly important~\cite{YL2002}. We observe that the relative difference between ground state energies calculated with PHF and CI decreases from $\approx$ 2\% up to $\approx$ 1\% as $\lambda$ increases from 2 to 8. This can be attributed to an increasing amount of correlations introduced by the PHF procedure when interactions grow. This trend was already pointed out for $N=2$ when spin and rotational symmetries were restored~\cite{YL2002}.

We close this section outlining the determination of the PHF ground state in the case $N=4$ with $\lambda=6$.
\begin{table}[htbp]
\begin{center}
\begin{tabular}[htbp]{|c||l|l|}
\hline
UHF state&$E_{i}^{S_{z}}$&$\bar{E}_{i,\bar{L}_{i},\bar{S}_{i}}^{S_{z}}$ ($\bar{L}_{i},\bar{S}_{i}$)\\
\hline \hline
$|\Psi_{1}^{S_{z}=2}\rangle$&24.139&24.022 (L=2, S=2)\\
\hline
$|\Psi_{1}^{S_{z}=1}\rangle$&24.169&23.891 (L=0,S=1)\\
\hline
$|\Psi_{1}^{S_{z}=0}\rangle$&24.170&23.884 (L=0, S=0)\\
\hline
$|\Psi_{2}^{S_{z}=0}\rangle$&24.200&23.880 (L=0, S=1)\\
\hline
\end{tabular}
\end{center}
\caption{Determination of the PHF ground state in the case of $N=4$ for $\lambda=6$ (see table~\ref{table2}). The UHF states $|\Psi_{i}^{S_{z}}\rangle$ considered are shown in column 1, their energy is shown in column 2. Column 3 shows the minimum projected energy $\bar{E}_{i,\bar{L}_{i},\bar{S}_{i}}^{S_{z}}$ for each UHF state. Energies in units $\omega_{0}$.}
\label{table3}
\end{table}
Table~\ref{table3} shows the lower energies (column 2) of different UHF states (column 1) found for $S_{z}=0,1,2$. The UHF ground state is $|\Psi_{1}^{S_{z}=2}\rangle$, with energy $E_{1}^{S_{z}=2}=24.139\ \omega_{0}$. The other shown states, however, are very close in energy. This is the typical case where much care must be taken when performing the PHF procedure: projecting the UHF ground state (first line) does not give rise to the PHF ground state (see column 3). On the other hand $|\Psi_{2}^{S_{z}=0}\rangle$ (fourth line), with the highest UHF energy among the listed states, yields the PHF ground state. We have also checked that UHF states with higher energies (not shown in the table) do not yield a better PHF ground state energy.
%
%
%
%
\subsection{Magnetic field}
\label{withfield}
For non--zero magnetic field, the projection procedure provides physical features that are not correctly reproduced by the
UHF approximation. In the following, we assume standard parameters for GaAs: $m^{*}=0.067 m_{e}$, $\varepsilon=12.4$ and $g^{*}=-0.44$. We start with the $N=2$ case with $\omega_{0}=3.37 \ {\rm meV}$ as a
confinement energy, in order to compare our results with the available ED
data~\cite{pfannkuche}.
\begin{figure}
\begin{center}
\includegraphics[width=12cm,keepaspectratio]{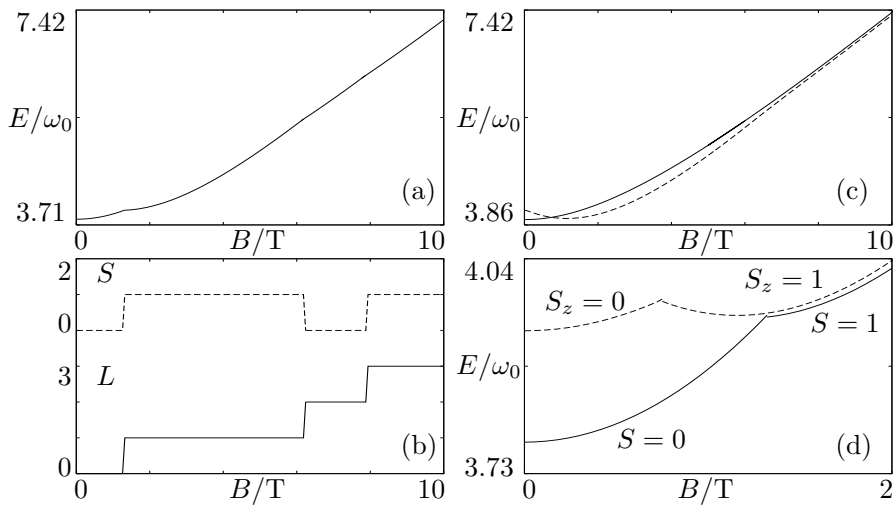}
\caption{PHF and UHF solutions for a quantum dot with $N=2$ electrons. (a) PHF ground
  state energy $E$ and (b) projected ground state total spin (dashed) and angular momentum
  (solid) as a function of the magnetic field $B$. (c) UHF energy $E$
  for the $S_{z}=0$ (solid) and $S_{z}=1$ (dashed) states as a
  function of $B$. (d) Comparison of PHF (solid) and UHF (dashed line) ground states in the $0\ {\rm T}\leq B\leq2\ {\rm T}$ field range. Corresponding spin quantum numbers are included. Parameters: $m^{*}=0.067 m_{e}$, $\varepsilon=12.4$, $g^{*}=-0.44$ and $\omega_{0}=3.37 \ {\rm meV}$.}
\label{fig1}
\end{center}
\end{figure} 
Figure~\ref{fig1}(a) shows the ground state energy of the PHF ground
state. Correspondingly, Fig.~\ref{fig1}(b) shows $L$ (solid) and $S$
(dashed) quantum numbers. With increasing $B$, an alternating sequence of singlet and
triplet states is observed as $L$ increases with unitary steps. PHF results compare
well with those obtained by means of more accurate methods~\cite{merkt,pfannkuche,dineykhan}. They predict both the same sequence of increasing $L$ and of singlet-triplet transitions.
Quantitatively, the crossover fields are not perfectly reproduced. For
instance the first singlet-triplet transition occurs at $B\approx 2 \ 
{\rm T}$~\cite{pfannkuche}, while we find $B=1.38 \ {\rm T}$.

The sequence of transitions between states with different total spins shown by the PHF solution
cannot be obtained using the UHF approximation. Fig.~\ref{fig1}(c) displays the UHF energies for $S_{z}=0$ (solid) and $S_{z}=1$ (dashed): only the $S_{z}=0\to 1$ transition at $B=0.75 \ {\rm T}$ can be found in the whole magnetic field range, in contrast to the PHF data. Fig.~\ref{fig1}(d) shows a comparison of the UHF and PHF ground states in the $0\ {\rm T}\leq B\leq2\ {\rm T}$ region. Clearly, at low magnetic fields the PHF singlet solution has a better energy. In addition, it is hard to relate the $S_{z}$ transition occurring in the UHF calculation with the genuine singlet--triplet transition exhibited by PHF. 
\begin{figure}
\begin{center}
\includegraphics[width=12cm,keepaspectratio]{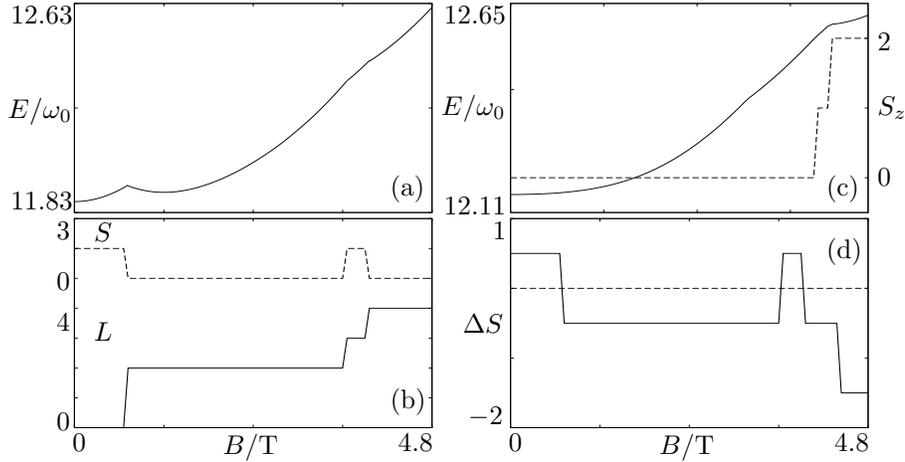}
\caption{Projected solutions for a quantum dot 
  with $N=3,4$ electrons. (a) Projected
  ground state energy $E$ (units meV) and (b) projected ground state
  total spin (dashed) and angular momentum (solid) for
  $N=4$, as a function of the magnetic field $B$. (c) UHF ground state energy (solid) and $S_{z}$ (dashed) for $N=4$ as a function of $B$. The inset shows the narrow transition of the UHF ground state to $S_{z}=1$. (d) Total
  spin difference between $N=4$ and $N=3$ PHF ground states, $\Delta
  S=S(4)-S(3)$, as a function of the magnetic field $B$.
  Dashed: $\Delta =0$. Parameters: $m^{*}=0.067 m_{e}$, $\varepsilon=12.4$, $g^{*}=-0.44$ and $\omega_{0}=6 \ {\rm meV}$.}
\label{fig2}
\end{center}
\end{figure} 

For discussing a quantum dot with four electrons, we choose a
confinement energy $\omega_{0}=6 \ {\rm meV}$ in order to be able to
compare our results with the ED calculations reported
earlier~\cite{maksymB}. The PHF ground state energy (Fig.~\ref{fig2}(a)),
shows kinks associated with spin and angular momentum transitions
(Fig.~\ref{fig2}(b)). We find a sequence of transitions which agrees with those reported
earlier~\cite{maksymB}: a low field triplet--singlet transition obtained with PHF occurs at $B\approx 0.7\ {\rm T}$, higher than the value $B\approx 0.5\ {\rm T}$ predicted by ED~\cite{maksymB}. Other PHF transitions are shifted to smaller values, similarly to the singlet--triplet transitions for $N=2$. As a comparison, we show (Fig.~\ref{fig2}(c)) the UHF ground state energy (solid line) and $S_{z}$ (dashed line) for $N=4$ in the same magnetic field range. Apart from the higher energy estimate provided by UHF, we observe that the spin transitions scenario is very different w.r.t. the one predicted by PHF. The UHF ground state has $S_{z}=0$ up to $B\approx 3.8\ {\rm T}$, in contrast with the above mentioned triplet--singlet transition obtained with PHF for $B\approx 0.7\ {\rm T}$. Subsequently, a narrow transition to $S_{z}=1$ is found. Then, for $B\gtrsim 4.2\ {\rm T}$ the UHF ground state is fully polarized. The sharp contrast with the PHF results confirms that the UHF method is not reliable to qualitatively predict the spin properties of the system, as we already observed in the $N=2$ case.

One of the important predictions of the results in~\cite{maksymB} is a
spin blockade~\cite{weinman} that in the ED calculation occurs at $4.96 \ {\rm T}<B<5.18 \ 
{\rm T}$, due to a violation of the total spin selection rules. In
Fig.~\ref{fig2}(d) we have plotted $\Delta S=S(4)-S(3)$ calculated by
using the PHF ground states for $N=4,3$ electrons. For $4.38 \ {\rm
  T}< B <4.8 \ {\rm T}$ we indeed observe $\Delta S=-3/2$ in analogy with the above mentioned results.

%
%
%
%
\section{Conclusions}
\label{conclusions}
We have applied the UHF method to quantum dots with up to four
electrons in the presence of a magnetic field, and for varying
strength of the interaction. We have used a systematic projection
approach for simultaneously restoring the total spin and the
rotational symmetries of the UHF wave functions. The projected wave functions are superpositions of many different
Slater determinants. Thus, they contain important correlations, missed by UHF solutions. We found that the energies of the ground states
obtained by PHF are systematically lower than those of the UHF ground states,
although still higher than the ones found with other methods such as ED, CI or QMC. The PHF provides ground
states with total spins and angular momenta in qualitative
agreement with the ``exact'' results. The data in table~\ref{table2} show that by means of PHF, important correlations are introduced for increasing $\lambda$. This is signalled also by the tendency of the PHF method to reproduce the CI results~\cite{rontani} with increasing accuracy.

We recover Hund's first rule for four electrons at zero magnetic
field.  With increasing magnetic field, crossovers between ground
states with different quantum numbers are found for fixed electron
number that are not reproduced by the UHF approximation. These are
consistent with the ones obtained with the ``exact'' techniques. We
have confirmed the presence of a spin blockade due to a spin mismatch
in the ground states of three and four electrons. 

We conclude that the PHF approach is an important technique when
dealing with ground state properties like total spin and angular
momentum of correlated electrons in quantum dots.  In view of what has
been discussed in Sec. \ref{secproj}, we expect that PHF will be very
useful to obtain interesting results for quantum dots with larger
electron numbers that, so far, have not been accessible by other
methods, especially in the presence of a magnetic field.

Such quantum dots have recently been experimentally
investigated \cite{RHCSK2006}. In particular, we expect that the wave
function obtained by PHF will be useful to obtain transition
matrix elements that are important for understanding the transport behavior.

\ack{This work has been initiated during a stay of U. De G. at I.
  Institut f\"ur Physik of the Universit\"at Hamburg within a Marie
  Curie collaboration of the EU (Contract No.  MCRTN-CT2003-504574)
  and a stay of F. C. at the International University Bremen. The work has been supported by the
  Italian MIUR via PRIN05, and by SFB 508 ``Quantenmaterialien'' of
  the Universit\"at Hamburg.}

\appendix

\section{The projection method for three particles}
\label{toy}In order to illustrate the procedure leading to the evaluation of (\ref{energyp}), we discuss here the case
 of three electrons. The possible UHF solutions have $S_{z}=\pm 1/2$ or $S_{z}=\pm 3/2$. For the latter case no spin projection is required since it is a pure $S=3/2$ state, so that only angular momentum projection has to be performed. Therefore, we concentrate here on the $S_z=1/2$ case. The UHF wavefunction reads from eq. (\ref{det}) 
\begin{equation}
\left|\Psi^{S_{z}=1/2}\right>=\left|T_0\right>=\frac{1}{\sqrt{6}}
{\rm det}\{a_{1}{\alpha},a_{2}{\alpha},b_{1}{\beta}\}
\end{equation} 
(the case $S_{z}=-1/2$ is obtained by interchanging all $\alpha$ and
$\beta$ spinors, without any conceptual difference with respect to the
following discussion). This state does not have defined total spin. It
is a superposition of the doublet $S=1/2$ and the quadruplet $S=3/2$
states.  The spin projection allows to select the total spin part we
are interested in. To do so, we generate all the possible {\em
shuffled} slater determinants $|T_q\rangle$ starting from
$|T_0\rangle$ and weight them by the corresponding Sanibel
coefficients (\ref{sanibel}), from eq. (\ref{project1})

\begin{equation}\label{san1}
\hat{P}_{S=1/2}^{S_z=1/2}\left|\Psi^{S_{z}=1/2}\right>= \frac{2}{3}|T_0\rangle-\frac{1}{3}|T_1\rangle\, ,
\end{equation}

\begin{equation}\label{san2}
\hat{P}_{S=3/2}^{S_z=1/2}\left|\Psi^{S_{z}=1/2}\right>= \frac{1}{3}|T_0\rangle+\frac{1}{3}|T_1\rangle\, ,
\end{equation}
where
\begin{equation}
|T_1\rangle=|T_1^{(1)}\rangle+|T_1^{(2)}\rangle =\frac{1}{\sqrt{6}}({\rm det}\{a_{1}{\alpha},a_{2}{\beta},b_{1}{\alpha}\}
+{\rm det}\{a_{1}{\beta},a_{2}{\alpha},b_{1}{\alpha}\})\, .
\end{equation}
In both (\ref{san1}) and (\ref{san2}), the $|T_0\rangle$ and
$|T_1\rangle$ terms correspond to all the possible ways to exchange
$q=0$ and $q=1$ pairs of spinors ($\alpha$, $\beta$) in the UHF
determinant.  Since states with $S=1/2$ and $S=3/2$ constitute the
only spin configurations of three electrons with $S_z=1/2$, summing up
(\ref{san1}) and (\ref{san2}) yields the original determinant $\mid
\Psi ^{S_{z}=1/2}\rangle$, consistently with the identity satisfied by
the sum of the projection operators.

For simplicity, in the following we consider the $S=1/2$ case only.
In order to obtain (\ref{project2}) we apply $\hat{P}_L$ on
(\ref{san1}). The action of the rotation generator
$\exp(i\hat{L}\gamma)$ on the determinants composing (\ref{san1})
affects the spatial part of the orbitals only: $a_i \to a_i(\gamma)$
and $b_i \to b_i(\gamma)$.  Where $a_i(\gamma)$, $b_i(\gamma)$ are the
UHF orbitals rotated by an angle $\gamma$ over the $z$ axis.
Employing (\ref{energyp}) and (\ref{project2}) the projected energy
can be recast into
\begin{equation}
E^{S_z=1/2}_{L,S=1/2}=\frac{N_{L,S=1/2}^{S_z=1/2}}{D_{L,S=1/2}^{S_z=1/2}}
\end{equation}
with
\begin{equation}\label{num}
N^{S_z=1/2}_{L,S=1/2}\!\!=\!\!\!\int_{0}^{2\pi}\frac{{\rm d}\gamma}{2\pi}\ e^{-iL\gamma}
\left[\frac{2}{3}\langle T_0|\hat{H}_0+\hat{V}|T_{0}(\gamma)\rangle-\frac{1}{3}\langle T_0|\hat{H}_0+\hat{V}|T_{1}(\gamma)\rangle\right]  
\end{equation}
\begin{equation}\label{den}
D^{S_z=1/2}_{L,S=1/2}\!\!=\!\!\!\int_{0}^{2\pi}\frac{{\rm d}\gamma}{2\pi}\ e^{-iL\gamma}
\left[\frac{2}{3}\langle T_0|T_{0}(\gamma)\rangle-\frac{1}{3}\langle T_0|T_{1}(\gamma)\rangle\right]\, .  
\end{equation}

The evaluation of (\ref{num}) and (\ref{den}) is done using the
standard theorems for many body wave functions \cite{loewdin2}. The
overlap terms in the denominator are $\langle
T_0|T_{0}(\gamma)\rangle={\rm det}\{{\bf d}^{(0)}(\gamma)\}$ and
$\langle T_0|T_{1}(\gamma)\rangle=\langle
T_0|T_{1}^{(1)}(\gamma)\rangle+\langle T_0|T_{1}^{(2)}(\gamma)\rangle
={\rm det}\{{\bf d}^{(1)}(\gamma)\}+{\rm det}\{{\bf
d}^{(2)}(\gamma)\}$ respectively, with
\begin{equation}
{\bf d}^{(0)}(\gamma)=
\left( \begin{array}{ccc}
\langle a_1|a_1(\gamma)\rangle & \langle a_1|a_2(\gamma)\rangle & 0\\
\langle a_2|a_1(\gamma)\rangle & \langle a_2|a_2(\gamma)\rangle &0\\
0                       &       0                 &\langle b_1|b_1(\gamma)\rangle
\end{array}\right)\, ,
\end{equation}
\begin{equation}
{\bf d}^{(1)}(\gamma)=
\left( \begin{array}{ccc}
\langle a_1|a_1(\gamma)\rangle & 0&\langle a_1|b_1(\gamma)\rangle \\
\langle a_2|a_1(\gamma)\rangle & 0&\langle a_2|b_1(\gamma)\rangle \\
0                       &   \langle b_1|a_2(\gamma)\rangle &0
\end{array}\right)\, , 
\end{equation}
\begin{equation}
{\bf d}^{(2)}(\gamma)=
\left( \begin{array}{ccc}
0&\langle a_1|a_2(\gamma)\rangle & \langle a_1|b_1(\gamma)\rangle \\
0&\langle a_2|a_2(\gamma)\rangle & \langle a_2|b_1(\gamma)\rangle \\
\langle b_1|a_1(\gamma)\rangle &0 &0
\end{array}\right)\, .
\end{equation}
The evaluation of the numerator, requiring the calculation of one--
and two--body operator matrix elements, is more involved. The single
particle part is
\begin{equation}
\langle T_0|\hat{H}_0|T_0(\gamma)\rangle+\langle T_0|\hat{H}_0|T_1(\gamma)\rangle=
\sum_{i=0}^2\sum_{k,l=1}^{3}h^{(i)}_{kl}(\gamma)d^{(i)}_{(k|l)}(\gamma)
\end{equation}
where $d^{(i)}_{(k|l)}(\gamma)$ is the $(k,l)$ entry of the first
order cofactor of ${\bf d}^{(i)}(\gamma)$ and
$h^{(i)}_{kl}(\gamma)=\langle u_k|\hat{H}_0|u_l(\gamma)\rangle M_{k
l}^{(i)}$. To save space in the above expression we have introduced
the notation $u_i\in \{a_1,a_2,b_1\}$ and the matrices
\begin{equation}
{\bf M}^{(1)}\!=\!
\left( \begin{array}{ccc}
1&1&0 \\
1&1&0\\
0&0&1
\end{array}\right)
\, ,\,
{\bf M}^{(2)}\!=\!
\left( \begin{array}{ccc}
1&0&1 \\
1&0&1\\
0&1&0
\end{array}\right)
\, , \,
{\bf M}^{(3)}\!=\!
\left( \begin{array}{ccc}
0&1&1 \\
0&1&1\\
1&0&0
\end{array}\right)
\, .
\end{equation} 
In a similar way the two body operator matrix element is
\begin{equation}
\langle T_0|\hat{V}|T_0(\gamma)\rangle+\langle T_0|\hat{V}|T_1(\gamma)\rangle=
\frac{1}{2}\sum_{i=0}^2\sum_{k_1,k_2,l_1,l_2=1}^{3}v^{(i)}_{k_1k_2l_1l_2}(\gamma)d^{(i)}_{(k_1k_2|l_1l_2)}(\gamma)
\end{equation}
where $d^{(i)}_{(k_1k_2|l_1l_2)}(\gamma)$ is the $(k_1,k_2,l_1,l_2)$
entry of the second order cofactor of ${\bf d}^{(i)}(\gamma)$ and
$v^{(i)}_{k_1k_2l_1l_2}(\gamma)=\langle
u_{k_1}u_{k_2}|\hat{V}|u_{l_1}(\gamma)u_{l_2}(\gamma)\rangle
M^{(i)}_{k_1 l_1}M^{(i)}_{k_2 l_2}$. The evaluation of these terms is
lengthy but straightforward.  Finally, the integrals in (\ref{num})
and (\ref{den}) are numerically evaluated by means of fast Fourier
transform once the spin projection has been performed.
\newline

\end{document}